\documentclass[runningheads]{llncs}
\usepackage[T1]{fontenc}
\usepackage{graphicx}
\usepackage{amsmath,amssymb} 
\usepackage{color}

\usepackage{tikz}
\usepackage{comment}
\usepackage{amsmath,amssymb} 
\usepackage{color}

\usepackage{booktabs}
\usepackage{multirow}
\usepackage{adjustbox}
\newcommand{\model}{RDRN}
\newcommand{\modelfullnametitle}{{Recursively Defined Residual Network}}
\newcommand{\modelfullname}{{recursively defined residual network}}
\newcommand{\blockname}{RDRB}
\newcommand{\blockfullname}{{recursively defined residual block}}
\newcommand{\blockfullnametitle}{{Recursively Defined Residual Block}}

\begin{document}
\title{\model: \modelfullnametitle{}  for Image Super-Resolution}
%
%
\author{Alexander Panaetov \and
Karim Elhadji~Daou\and
Igor Samenko \and
Evgeny Tetin \and
Ilya Ivanov} 

\authorrunning{A. Panaetov et al.}
%
\institute{Huawei, Moscow Research Center, Russia \email{\{panaetov.alexander1, karim.daou, samenko.igor, evgeny.tetin, ivanov.ilya1\}@huawei.com}}
\maketitle              
\begin{abstract}
	Deep convolutional neural networks (CNNs) have obtained remarkable performance in single image super-resolution (SISR). 
However, very deep networks can suffer from training difficulty and hardly achieve further performance gain. 
There are two main trends to solve that problem: improving the network architecture for better propagation of features through large number of layers and designing an attention mechanism for selecting most informative features. Recent SISR solutions propose advanced attention and self-attention mechanisms. However, constructing a network to use an attention block in the most efficient way is a challenging problem. To address this issue, we propose a general \blockfullname{} (\blockname) for better feature extraction and propagation through network layers. 
Based on \blockname{} we designed \modelfullname{} (\model), a novel network architecture which utilizes attention blocks efficiently. Extensive experiments show that the proposed model achieves state-of-the-art results on several popular super-resolution benchmarks and outperforms previous methods by \textbf{up to 0.43 dB.}

\end{abstract}
\begin{figure}[t]
	\centering
	\includegraphics[width=0.98\linewidth]{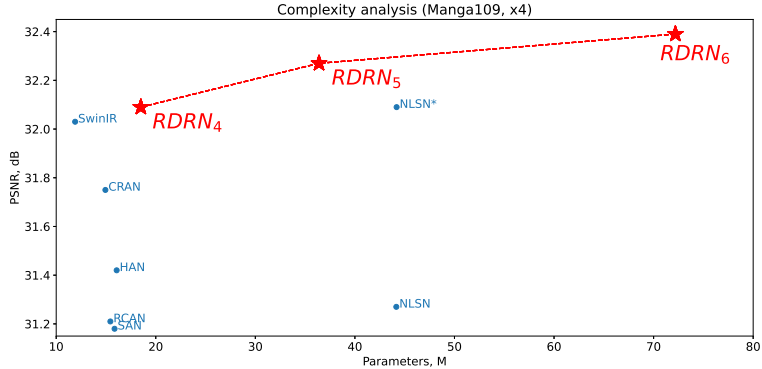}
	\caption{Number of parameters and performance on Manga109 with upscale factor $\times 4$ (BI model) }
	\label{fig:my}
\end{figure}

\section{Introduction}
\begin{figure}[t]\footnotesize
	\begin{center}
		\includegraphics[width = \textwidth]{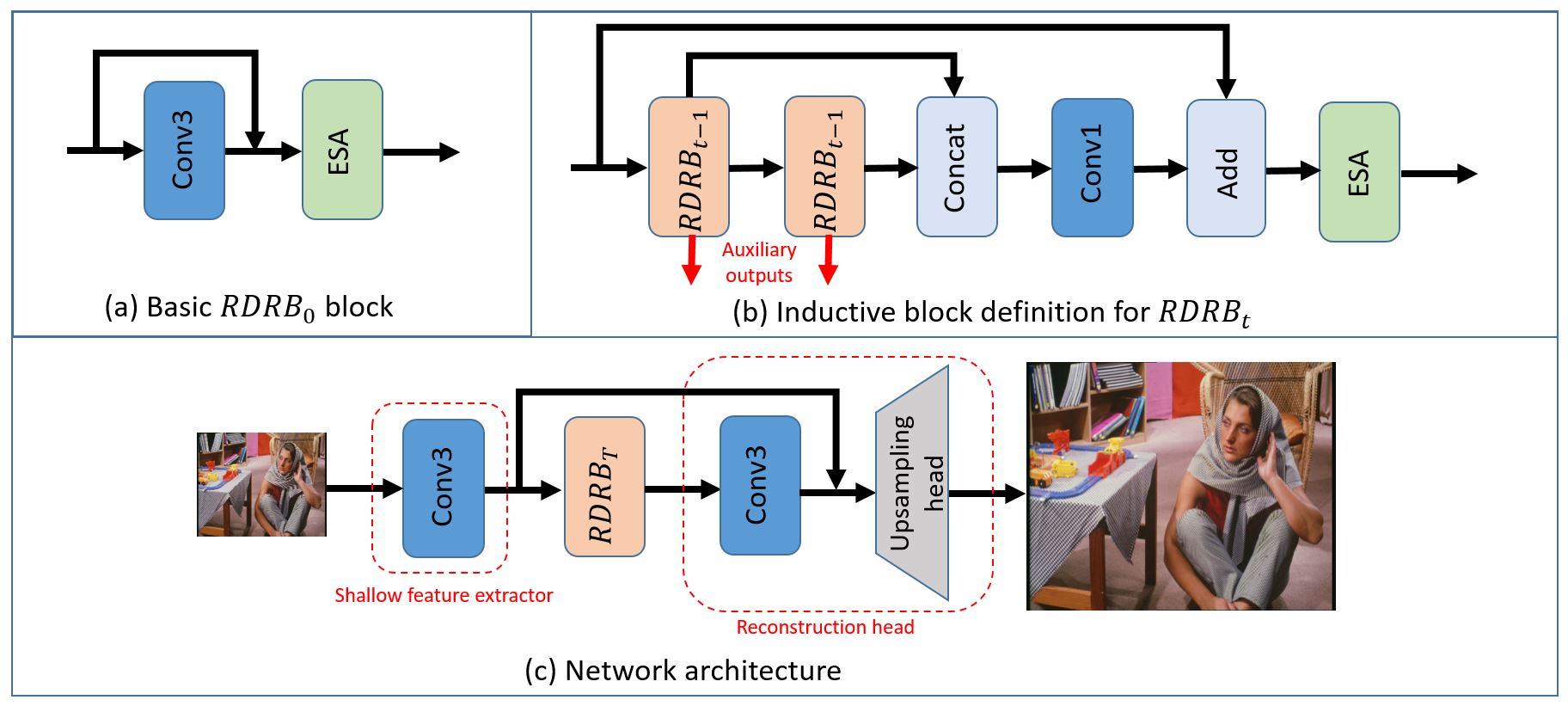}
	\end{center}

	\caption{ Architecture of the proposed \blockfullname{} (\blockname) and network (\model). \blockname{} is defined in a recurrent manner. (a) We define basic block $\blockname_0$ as a convolution layer followed by ESA block from \cite{RFANet}.
	(b) Each subsequent block $\blockname_t$ can be defined using the architecture of previous block $\blockname_{t-1}$ according to the scheme. (c) Full network architecture 
	} 
	\label{fig-main}
\end{figure}

The main purpose of super-resolution (SR) is to reconstruct high-resolution image (HR) from given low-resolution counterpart (LR).
SR is an ill-posed problem since the mapping between LR and HR images is ambiguous (one-to-many). Recovering missing details is a challenging task, especially for a high upscale factor. Despite being a difficult problem, SR plays an important role in various image processing tasks with applications in face recognition, medical imaging, surveillance, digital zoom, etc. While many existing SR methods reconstruct HR image from several LR images, in this paper we focus on single image super-resolution (SISR). 

In recent years, convolutional neural networks (CNNs) have achieved remarkable results in many computer vision tasks, including SISR.
Deep CNNs have shown improvement over the traditional algorithms. Network depth in existing solutions has been significantly increased from three layers in SRCNN \cite{dong2015image} to more than 400 in recent works \cite{zhang2018image,HAN}. 
However, very deep networks can suffer from training difficulties and hardly achieve any extra performance gain. 
A further increase in CNN depth does not lead to an improvement in quality and makes them unsuitable for various applications. 
The difficulty of training can be explained by the fact that network is not able to efficiently use information from intermediate layers. This issue can be partially solved using residual learning \cite{he2016deep}. 
Combining features from different layers through skip connections is a fruitful idea in SISR. Additional connections along the network's depth could help to learn more powerful feature representations, making training more stable and accelerating convergence.

Another approach to address the training difficulty is related to the mechanism of attention. Recently, this direction has become very popular and profitable for SISR. The intuition behind attention is a simulation of the human vision system, which can focus on the most informative parts of an image and ignore the irrelevant information. Recent works show that attention can effectively reduce the width and depth of a network while maintaining comparable or better performance due to enhanced discriminative learning ability \cite{zhang2018image}.

In this paper, we combine both approaches. The design of \blockfullname{} (\blockname) is shown in Figure \ref{fig-main}. It consists of two parts: basic block (Fig. \ref{fig-main}-a) and recursive block (Fig. \ref{fig-main}-b). We have found that enhanced spatial attention (ESA) introduced in \cite{RFANet} is very effective for the super-resolution task, and we take advantage of its benefits even more than in the original paper. We include ESA in the basic block, which is repeated in the final architecture multiple times. Compared to previous work, our \blockname{} contains more connections between intermediate layers. It combines hierarchical cues along the network depth to obtain richer feature representations. Experiments show that the effect of the proposed \blockname{} is more visible for lower upscale factors ($\times2$, $\times3$). For the upscale factor $\times 2$, the \blockname{}-based model outperforms all recent solutions without any bells and whistles. It can be explained by the recurrent nature of \blockname. Shallow features are propagated to all levels of \blockname{} via long skip connections from input. For lower upscale factors, they contain more relevant information, as fewer details will be missed compared to higher upscale factors. Finally, to further improve \blockname{}, especially for large upscale factors, we insert non-local sparse attention \cite{NLSN} into the block.

Based on \blockname{} we design \modelfullname{} (\model) as shown in Figure \ref{fig-main}-c. Following \cite{adadm}, we add batch normalization (BN) and apply adaptive deviation modulator (AdaDM) to the final model. For training, we use intermediate supervision (IS) loss, which improves convergence and allows to simultaneously train several models of different computational complexity without a large overhead.

The proposed \model{} shows superior performance on the most popular SR benchmarks. Our model produces better visual quality, recovers more details and outperforms current state-of-the-art solutions with a significant margin of up to 0.43 dB.

In summary, the main contributions of our paper are:
\begin{itemize}
	\item We propose a novel recursive scheme for block architecture definition. Using that scheme we build  a general \blockfullname{} (\blockname)  for more accurate image SR.
	\item Based on \blockname{} we design a novel network architecture (\model). Extensive experiments on public datasets demonstrate that the proposed model outperforms current state-of-the-art SR methods.	
	\item  We introduce intermediate supervision loss. Training with IS helps to obtain additional performance gain and allows to simultaneously train several models of different computational complexity.
\end{itemize}

\section{Related Work}

Super-resolution algorithms can be categorized into two types: traditional and deep learning based methods. In this section, we will focus on the second category as the most successful in computer vision.

Dong et al.\cite{dong2015image} proposed the first three-layer convolutional neural network (SRCNN) to learn the mapping from LR to HR directly. This pioneering work achieved superior performance against the previous traditional methods.

Following this work, many networks achieved better performance using deeper and wider architectures \cite{kim2016accurate,kim2016deeply}. 
SRCNN used interpolated image as input, however, it is more efficient to upscale the feature maps at the end of the network. To address this issue, Shi et al. proposed ESPCN \cite{shi2016real} with a sub-pixel layer, which is widely used in modern SISR networks. As in the case of the classification task, a further increase in depth and width of plain architecture leads to quality degradation. However, residual blocks \cite{he2016deep} and dense connections \cite{huang2017densely} allow to train more powerful networks, and Lin et al. proposed a very deep and wide EDSR \cite{lim2017enhanced} based on a modified residual block. 
Dense connections were used in RDN \cite{zhang2018residual} to utilize hierarchical features from all convolutional layers.
  
Following \cite{lim2017enhanced}, most of the recent SISR works do not use batch normalization (BN) \cite{ioffe2015batch}, as it harms network's performance. However, Liu et al. \cite{adadm} showed that normalization layers reduce the standard deviation of feature pixels (the main reason for quality degradation) and proposed adaptive deviation modulator (AdaDM) to solve that issue. AdaDM can successfully enable BN layers, significantly improving performance and allowing to train larger models.

Attention mechanism is used as a simulation of the human vision system that focuses on the most informative parts of an image, ignores irrelevant information and enhances discriminative learning ability. Recently, attention has been successfully applied to the SISR problem, and attention-based methods have shown superiority over pure CNN solutions. Zhang et al. \cite{zhang2018image} proposed channel attention (CA) to adaptively rescale each feature channel-wise by modeling the interdependencies across feature channels. Such CA mechanism improves the representational ability of residual channel attention network (RCAN).

Liu et al. \cite{RFANet} designed lightweight and powerful enhanced spatial attention (ESA). 
Dai et al. \cite{dai2019second} introduced second-order attention network (SAN) to adaptively refine features using second-order feature statistics.
Niu et al.\cite{HAN} presented a new holistic attention network (HAN), which consists of a layer attention module (LAM) and a channel-spatial attention module (CSAM) and models the holistic interdependencies among layers, channels, and positions.
Zhang et. al. \cite{CRAN} proposed a context reasoning attention network (CRAN) that can adaptively modulate the convolution kernel according to the global context enhanced by semantic reasoning.
  


Attention could be considered as an additional tool to improve the expressiveness and convergence of CNNs.
An alternative is the self-attention mechanism \cite{attention}, which was introduced for natural language processing models. Recent work has shown that even pure transformers can achieve SOTA results on computer vision tasks \cite{vit}. However, hybrid models that contain both convolution and self-attention show outstanding results as they integrate the advantages of both approaches. Following this new trend, Liang et al \cite{swinir} introduced the SwinIR architecture based on Swin Transformer \cite{swin}. SwinIR has improved upon all previous methods by a significant margin.


\section{Proposed Method}
In this section, we first present the overview of \model{}. Then we give a detailed description of the proposed \blockname. We proceed with outlining the technical part of training, including the loss function and implementation details. Finally, we provide arguments about the advantages of the proposed method and differences from other approaches.

\subsection{Network Architecture}
As shown in Figure \ref{fig-main}, \model{} can be divided into three parts: shallow feature extractor, \blockname{} deep feature extractor and reconstruction head. Let us denote $I_{LR}$ and $I_{SR}$ the low resolution input and the output of \model. Following \cite{zhang2018image,dai2019second,HAN}, we use only one convolutional layer to extract the shallow features $F_{SF}$ from LR input:
\begin{equation}
\mathit{F_{SF}} =  H_{SF}(I_{LR}),
\end{equation}
where $ H_{SF}$ denotes the convolution operation. The extracted shallow feature map $F_{SF}$ is used as input to the deep feature extractor:

\begin{equation}
\mathit{F^{T}_{DF}} =  H^{T}_{\blockname}(F_{SF}),
\end{equation}
where $T$ is the recursion depth and $ H^{T}_{\blockname}$ stands for \blockname{} feature extraction model, which will be introduced in the next subsection. \blockname{} is the core of the proposed network. Finally, the extracted deep features $F_{DF}$ are combined with the shallow features to stabilize training, after which they are processed by the reconstruction module:
\begin{equation}
\mathit{I^{T}_{SR}}=H_{rec}(F_{SF}, F^{T}_{DF}),
\end{equation}
where $H_{rec}$ denotes the reconstruction head that consists of a convolutional layer and a sub-pixel layer \cite{shi2016real}. The long skip connection propagates low-frequency information directly to the reconstruction module, which can help the deep feature extractor to focus on the extraction of high-frequency information \cite{swinir}.

There are several choices for loss function to optimize the model, such as $\ell_1$, $\ell_2$, perceptual, 
adversarial loss. We found that for the proposed method, $\ell_1$ loss is the most suitable one, and we minimize the following loss function:

\begin{equation}
\mathit{L^T(\Theta)} = \frac{1}{m}\sum_{i=1}^{m}\big\|H^T_{\model}(I^{i}_{LR})-I^{i}_{HR}\big\|_{1}= \frac{1}{m}\sum_{i=1}^{m}\big\|I^{T,i}_{SR}-I^{i}_{HR}\big\|_{1},
\end{equation}
where $H^T_{\model}$, $ \Theta $, and $m$ denote the function of the proposed \model{}, the set of learned parameters, and the number of training pairs, respectively. Following \cite{RFDN} we fine-tune the model using $\ell_2$ loss. More details are provided in the experiments section.

%
%
%

\subsection{\blockfullnametitle{} (\blockname)}
As discussed above, \blockfullname{} is the core of \model{}, and here we give the overall description of the proposed block architecture. 
The definition will be done in a recursive manner. First, we define basic block $\text{\blockname}_0$ as a convolutional layer followed by ESA block introduced in \cite{RFANet}:

\begin{equation}
F^0_{DF} = H^0_{\blockname}(F_{SF}) := ESA\big(conv_{3\times 3} (F_{SF}) + F_{SF}\big).
\end{equation}
We found that ESA mechanism is highly effective for the super-resolution task, and we take advantage of its benefits even more than in the original paper. We include ESA in the basic block, which is repeated in our architecture multiple times. 

Finally, we use induction to define $\text{\blockname}_t$ for any natural t:
\begin{equation}
\label{eq1}
F^t_{DF} =  H^t_{\blockname}(F_{SF}),
\end{equation}
\begin{equation}
\label{eq2}
F^{t,1}_{DF} =  H^t_{\blockname}(F^t_{DF}),
\end{equation}

\begin{equation}
 H^t_{\blockname}(F_{SF}) := ESA\bigg(F_{SF} + conv_{1 \times 1}\Big( concat\big(F^{t-1}_{DF}, F^{t-1, 1}_{DF})\big)\Big)\bigg).
\end{equation}
The scheme of building $\text{\blockname}_0$ and $\text{\blockname}_t$ is depicted in Figure \ref{fig-main}.
Same as in the basic block, ESA is included in $\text{\blockname}_t$. It's worth noticing that we don't share weights in any part ot the model. To avoid any misunderstanding about the definition of the proposed RDRB, we provide PyTorch implementation of the basic block and the recursive step in the supplementary. 

\subsection{Intermediate Supervision (IS)}
As shown in Figure \ref{fig-main}, for $t > 0$ each $\text{\blockname}_t$ contains two additional auxiliary outputs. 
Each of these intermediate outputs is paired with a reconstruction head and an additional loss.  The final loss function is a weighted sum of the original loss and all intermediate losses (due to recursive definition of the block we have $2^{T+1} -2$ additional loss terms):
\begin{equation}
\mathit{L(\Theta)} = w_0 \mathit{L^T(\Theta)} + \sum\limits_{i=1}^{2^{T+1} -2} w_{i} \mathit{L_i(\Theta)},
\end{equation}
where $w_{i}$ denotes the weight and $\mathit{L_i(\Theta)}$ is the loss based on the intermediate output.
The computational overhead for IS training is minimal, because the size and complexity of the added heads is much smaller than the size and complexity of \blockname. Using IS loss function gives two advantages. First, it allows to simultaneously train several models of different computational complexity using the output of intermediate reconstruction head. Second, as proved in the ablation study, training with IS enables us to achieve a performance gain.

\subsection{Implementation Details}
Here we specify the implementation details of \model. We use $\text{\blockname}_5$ in our final architecture. 
Following \cite{adadm}, we add batch normalization (BN) \cite{ioffe2015batch} and adaptive deviation modulator (AdaDM) to every $3 \times 3$ convolution. Finally, starting from recursion level 3 we add non-linear spatial attention block \cite{NLSN} after ESA. According to our experiments, both tricks improve the final score and allow us to outperform SwinIR \cite{swinir}. 

Experiments show that for IS training it is better to zero out the weights for losses based on auxiliary outputs from $\text{\blockname}_1$ and $\text{\blockname}_2$ and set the remaining weights to one. 

\subsection{Discussion} 
In our research we aim to follow the best practices from recent work.  The motivation is two-fold: to outperform current SoTA models and to design general method for architecture definitions which can be used by other researchers. We conducted a large number of experiments and devised the following recipe:

\begin{itemize}
	\item Following \cite{RFDN}, we use advanced two-stage training procedure.
	\item We apply batch normalization \cite{ioffe2015batch} and add AdaDM to the model as in \cite{adadm}.
	\item We have an attention mechanism both inside the basic block and between blocks.
\end{itemize}
We explored deformable convolution \cite{dconvmod} for image SR. In all experiments it improves the final score but doubles the training time, and we decided to postpone this direction for future research.

The main novelty of our work is the proposed architecture of \blockname{} and
\model. We present the concept of building a network that utilizes multiple
connections between different blocks. It is not limited to one specific architecture:
the basic block can be replaced with any other block and the merging of two
blocks can vary as well. The proposed architecture significantly differs
from the ones presented in prior works.

\textbf{Difference from Residual Feature Aggregation Network (RFANet).}

RFANet \cite{RFANet} utilizes skip-connection only on block level while blocks are connected consequently. The intuition behind \model{} is that combining of hierarchical cues along the network depth allows to get richer feature representations. Compared to RFANet, our architecture is able to combine features at deeper layers. In ablation study we have comparison of residual-block (RB) from RFANet and $\text{\blockname}_0$ under the same conditions. We also show in Table \ref{tab1} the benefits of our recursive way of building the network compared to plain connection of blocks in RFANet. 

\textbf{Difference from Non-Local Sparse Network (NLSN).} 
NLSN \cite{NLSN} is a recent state-of-the-art SR method. The model achieved remarkable results after it was updated with BN and AdaDM \cite{adadm}. The original architecture of NLSN is plain and made of residual blocks and non-local sparse attention (NLSA) blocks. The main difference is again coming from more effective way of merging information from different layers. Our smaller $\text{\model}_4$ model has similar performance with NLSN and 72\% less FLOPs (Table \ref{table:kysymys}).

\textbf{Difference from Deep Recursive Residual Network (DRRN).} 
The architecture of the recursive block (RB) from \cite{tai2017image} is defined by a recursive scheme, similarly to the proposed \blockname. The key differences between the two approaches are as follows. First, according to the definition of RB, the weights of the same sub-blocks are shared, while \blockname{} does not reuse weights. However, weights sharing can be an effective way to reduce the number of parameters, and we save this direction for future research. Second, DRRN contains sequences of RB blocks. In contrast, our model is based on one large \blockname{}. Our recursive definition helps to stack blocks together with additional skip connections, granting extra performance gain.

\section{Experiments}
%
In this section, we compare \model{} to the state-of-the-art algorithms on five benchmark datasets. We first give a detailed description of experiment settings. Then we analyze the contribution of the proposed neural network. Finally, we provide a quantitative and visual comparison with the recent state-of-the-art methods for the most popular degradation models.    
%
\subsection{Implementation and trainings details}
\noindent\textbf{Datasets and metrics.} We train our models on DF2K dataset, which combines DIV2K~\cite{timofte2017ntire} and Flickr2K together as in \cite{CRAN,swinir,adadm}.
For testing, we choose five standard datasets: Set5~\cite{bevilacqua2012low}, Set14~\cite{zeyde2010single}, B100~\cite{martin2001database}, Urban100~\cite{huang2015single}, and Manga109~\cite{matsui2017sketch}. 
We train and test \model{} for three degradation methods. Degraded data is obtained by bicubic interpolation (BI), blur-downscale (BD) and downscale-noise (DN) models from Matlab. We employ peak signal-to-noise ratio (PSNR) and structural similarity (SSIM) \cite{SSIM} to measure the quality of super-resolved images. All SR results are evaluated on Y channel after color space transformation from RGB to YCbCr. Metrics are calculated using Matlab.

\textbf{Training settings.} The proposed network architecture is implemented in PyTorch framework, and models are trained from scratch. 
In our network, patch size is set to $ 64 \times 64 $. We use Adam \cite{kingma2014adam} optimizer with a batch size of 48 (12 per each GPU). The initial learning rate is set to $ 10^{-4} $. After $7.5 \times 10^{5}$ iterations it is reduced by 2 times. Default values of $ \beta_{1}$ and $\beta_{2}$ are used, which are 0.9 and 0.999, respectively, and we set $ \epsilon=10^{-8} $.
During training we augment the images by randomly rotating $90^{\circ}$, $180^{\circ}$, $270^{\circ}$ and horizontal flipping.

For all the results reported in the paper, we train the network $9 \times 10^{5} $ iterations with $\ell_1$ loss function. After that we fine-tune the model $1.5 \times 10^{4}$ iterations with a smaller learning rate of $ 10^{-5}$ using MSE loss. Complete training procedure takes about two weeks on a server with 4 NVIDIA Tesla V100 GPUs.

\subsection{Ablation study}


\textbf{Impact of the architecture.}
To highlight the advantages of the proposed architecture, we compare \model{} and RFANet \cite{RFANet}. For comparison we use vanilla RFANet with 32 RFA blocks and 64 channels. Each RFA block consists of 4 residual blocks (RB). First, we demonstrate that adding ESA to our basic block $\text{\blockname}_0$ is beneficial. For that purpose we train RFANet($\text{\blockname}_0$), changing RB to $\text{\blockname}_0$. As shown in Table \ref{tab1}, our basic block gives a performance gain to RFANet. 
Second, we show that the proposed recurrent scheme of building a network is better than stacking RFA blocks. We train \model(RB) using RB instead of $\text{\blockname}_0$ to demonstrate that. Even when using the same basic block, \model(RB) still outperforms RFANet(RFA, RB). 
For fair comparison, we change the depth of both networks to keep the computation complexity for all models similar.
 
\begin{table}[!t]\scriptsize
	
	\tabcolsep 10pt
	\caption{Ablation study on the advantage of \blockname}
	
	\begin{center}\scriptsize{
			\begin{tabular}{cccccc}
				\toprule
				Model / PSNR &Set5 & Set14 & B100 & Urban100 & Manga100 \\
				\midrule
				RFANet(RB)              & 32.65	& 29.00 & 27.86 & 27.11 & 31.73   \\
			    \midrule
			    RFANet($\text{\blockname}_0$)  & 32.67  & 29.02 & 27.86 & 27.16 & 31.81  \\
			    							 & \bf +0.02  &  \bf +0.02 & 0.00 & \bf +0.05 &  \bf +0.08 \\
				\midrule
				\model(RB)                   & 32.73 & 29.02 &	27.86	 & 27.14 & 31.80 \\
				 							 & \bf +0.08  & \bf +0.02 & 0.00 &  \bf +0.03 & \bf +0.07 \\
		\bottomrule		  
		\end{tabular}}
		\label{tab1}
	\end{center} 
\end{table}

\textbf{Impact of Intermediate Supervision (IS).}
To show the impact of using IS during training, we train the same network with and without IS. Table \ref{tab-IS} demonstrates the effectiveness of the proposed loss function. Experiments show that IS provides a bigger gain for larger models. It corresponds with the intuition that it is harder to train a larger model using SGD, and IS helps to propagate gradients better. Using IS allows to simultaneously train several SISR models of different computational complexity.    

\begin{table}[!t]\scriptsize
	
	\tabcolsep 10pt
	\caption{Ablation study on the advantage of IS}
	
	\begin{center}\scriptsize{
			\begin{tabular}{cccccc}
				\toprule
			Model / PSNR	&Set5 & Set14 & B100 & Urban100 & Manga100 \\
				\midrule
				$\model_4$    & 32.68 & 29.00 & 27.84 & 27.07 & 31.77 \\
				$\model_4$-IS & 32.71 & 29.01 & 27.85 & 27.07 & 31.83 \\
				              &\bf +0.03  & \bf +0.01 & \bf +0.01 & 0.00 & \bf +0.06 \\
				\midrule
				$\model_5$    & 32.67 & 28.99 & 27.86 & 27.09 & 31.86 \\
				$\model_5$-IS & 32.73 & 29.05 & 27.88 & 27.19 & 31.97 \\
							  & \bf +0.06  &  \bf+0.06 & \bf +0.02 & \bf +0.10 & \bf +0.11 \\
				\bottomrule		  
		\end{tabular}}
		\label{tab-IS}
	\end{center} 
\end{table}

\subsection{Results with Bicubic (BI) Degradation Model}

We compare the proposed algorithm with the following 11 state-of-the-art methods: SRCNN~\cite{dong2014learning}, RDN~\cite{zhang2018residual}, RCAN~\cite{zhang2018image}, SAN~\cite{dai2019second}, NLSN \cite{NLSN}, DRLN \cite{anwar2019drln}, HAN \cite{HAN}, RFANet\cite{RFANet}, CRAN~\cite{CRAN}, SwinIR~\cite{swinir} and NLSN*~\cite{adadm}.
We take all results from the original papers.
Following~\cite{lim2017enhanced,dai2019second,zhang2018image,HAN,swinir}, we provide a self-ensemble model and denote it \model+.

\textbf{Quantitative results.} Table \ref{table-BI} reports the quantitative comparison of $\times 2$, $ \times 3$, $\times 4$ SR. Compared to the existing methods, \model{} scores best in all scales of the reconstructed test sets. Even without self-ensemble, our model outperforms other solutions at upscale factors $2\times$ and $3 \times$. For $4 \times$ SR, \model{} achieves the best PSNR values, however, SSIM score of competitors on several datasets is higher.
The effect of the proposed method is more significant for lower upscale factors. This can be explained by the \blockname{} design. Shallow features are propagated for all levels of \blockname{} using long skip connections from input. For lower upscale factors, shallow features contain more important information as less information will be missed compared to higher upscale factors.

\textbf{Visual results.} We give visual comparison of various competing methods on Urban100 dataset for $\times 2$ SR in Figure \ref{fig:visual_result_SRBIX4}. Our model obtains better visual quality and recovers a more detailed image. Most compared methods recover grid textures of buildings with blurring artifacts, while \model{} produces sharper images. The proposed method can maintain a regular structure in difficult cases where previous approaches fail. To further illustrate the analysis above, we show such cases for $\times 2$ SR in Figure \ref{fig:visual_result_SRBIX2}. The recovered details are more faithful to the ground truth.

\begin{table}[!ht]
	\scriptsize 
	\centering
	\caption{Quantitative results with BI degradation model. The best and second best results are highlighted in \textbf{bold} and \underline{underlined}}

\caption{Visual comparison for $4\times$ SR with BI model on Urban100 dataset.  
}
\label{fig:visual_result_SRBIX4}
\end{figure*}

\begin{figure*}[t]
	\scriptsize
	\centering
	%
	\end{adjustbox}
	\vspace{1mm}
	\\

	\end{tabular}
	\caption{Visual comparison for $2\times$ SR with BI model on Urban100 dataset.  
	}
	\label{fig:visual_result_SRBIX2}
\end{figure*}

\subsection{Results with Bicubic Blur-Downscale (BD) Degradation Model}
Following \cite{zhang2018residual,zhang2018image,RFANet,anwar2019drln,HAN,CRAN}, we provide results for blur-downscale degradation, where HR image is blurred by a 7x7 Gaussian kernel with standard deviation $\sigma=1.6$ and then downscaled by bicubic interpolation with scaling factor $\times 3$ .

\textbf{Quantitative results.} 
In Table \ref{tab-BD} we compare the proposed \model{} model with the following super-resolution methods:  SPMSR~\cite{peleg2014statistical}, SRCNN~\cite{dong2014learning}, FSRCNN~\cite{dong2016accelerating}, VDSR~\cite{kim2016accurate}, IRCNN~\cite{zhang2017learning}, SRMDNF~\cite{zhang2018learning}, RDN~\cite{zhang2018residual},  RCAN~\cite{zhang2018image}, SRFBN~\cite{li2019feedback}, SAN~\cite{dai2019second}, HAN~\cite{HAN}, FRANet \cite{RFANet} and CRAN \cite{CRAN}. As shown, our solution achieves consistently better performance than other methods even without self-ensemble (i.e. \model+).

\textbf{Visual results.}
We show visual comparison for $\times 3$ SR with BD degradation in Figure \ref{fig:visual_result_SRBDX3}. The proposed model can recover grid textures and stripes even under heavy blur conditions. In the provided examples, \model{} reconstructs all stripes in the correct direction. In contrast, compared models have problems with the stripes direction and blurred areas.

\begin{table}[t]
	\centering
	\scriptsize
	\caption{Quantitative results with BD degradation model. The best and second best results are highlighted in \textbf{bold} and \underline{underlined}}
	\begin{tabular}{|p{6.5em}|p{2.5em}|p{2.5em}|p{3em}|p{2.5em}|p{3em}|p{2.5em}|p{3em}|p{2.5em}|p{3em}|p{2.5em}|p{3em}|}
		
		\hline
		\multicolumn{1}{|c|}{\multirow{2}{*}{{ Method} }} & \multicolumn{1}{c|}{\multirow{2}{*}{Scale}} & \multicolumn{2}{c}{Set5} & \multicolumn{2}{c}{Set14} & \multicolumn{2}{c}{B100} & \multicolumn{2}{c}{Urban100} & \multicolumn{2}{c|}{ Manga109} \\
		\cline{3-12}  &  &  PSNR  & SSIM  & PSNR   & SSIM  & PSNR  & SSIM  & PSNR  & SSIM  & PSNR  & SSIM \\
		\hline
		
		Bicubic \newline{}SPMSR~\cite{peleg2014statistical} \newline{}SRCNN~\cite{dong2014learning}  \newline{}FSRCNN~\cite{dong2016accelerating} \newline{}VDSR~\cite{kim2016accurate} \newline{}IRCNN~\cite{zhang2017learning}  \newline{}SRMDNF~\cite{zhang2018learning} \newline{}RDN~\cite{zhang2018residual} \newline{}RCAN~\cite{zhang2018image} \newline{} SRFBN~\cite{li2019feedback} \newline{} SAN~\cite{dai2019second}  \newline{} HAN~\cite{HAN} \newline{} RFANet \cite{RFANet} \newline{} CRAN \cite{CRAN}  \newline{} \model (ours) \newline{} \model +(ours)	&
		$\times3$ \newline{}$\times3$ \newline{}$\times3$ \newline{}$\times3$ \newline{}$\times3$ \newline{}$\times3$ \newline{}$\times3$ \newline{}$\times3$ \newline{}$\times3$ \newline{} $\times3$ \newline{}$\times3$ \newline{}$\times3$ \newline{}$\times3$ \newline{}$\times3$ \newline{}$\times3$ \newline{}$\times3$ 
		& 28.78 \newline{}32.21 \newline{}32.05 \newline{}26.23 \newline{}33.25 \newline{}33.38 \newline{}34.01 \newline{}34.58 \newline{}34.70 \newline{}34.66 \newline{}34.75 \newline{}{34.76} \newline{}{34.77} \newline{}{34.90}  \newline{}\underline{35.07} \newline{}\textbf{35.12}&
		0.8308 \newline{}0.9001 \newline{}0.8944 \newline{}0.8124 \newline{}0.9150 \newline{}0.9182 \newline{}0.9242 \newline{}0.9280 \newline{}0.9288 \newline{}0.9283 \newline{}0.9290 \newline{}{0.9294} \newline{}{0.9292} \newline{}{0.9302}  \newline{}\underline{0.9317} \newline{}\textbf{0.9320} &
		26.38 \newline{}28.89 \newline{}28.80 \newline{}24.44 \newline{}29.46 \newline{}29.63 \newline{}30.11 \newline{}30.53 \newline{}30.63 \newline{}30.48 \newline{}30.68 \newline{}{30.70} \newline{}{30.68}  \newline{}{30.79}  \newline{}\underline{31.07} \newline{}\textbf{31.15} &
		0.7271 \newline{}0.8105 \newline{}0.8074 \newline{}0.7106 \newline{}0.8244 \newline{}0.8281 \newline{}0.8364 \newline{}0.8447 \newline{}0.8462 \newline{}0.8439 \newline{}0.8466 \newline{}{0.8475} \newline{}{0.8473} \newline{}{0.8485} \newline{}\underline{0.8524} \newline{}\textbf{0.8533}&
		26.33 \newline{}28.13 \newline{}28.13 \newline{}24.86 \newline{}28.57 \newline{}28.65 \newline{}28.98 \newline{}29.23 \newline{}29.32 \newline{}29.21 \newline{}29.33 \newline{}{29.34} \newline{}{29.34}  \newline{}{29.40}  \newline{}\underline{29.54} \newline{}\textbf{29.57} &
		0.6918 \newline{}0.7740 \newline{}0.7736 \newline{}0.6832 \newline{}0.7893 \newline{}0.7922 \newline{}0.8009 \newline{}0.8079 \newline{}0.8093 \newline{}0.8069 \newline{}0.8101 \newline{}{0.8106} \newline{}{0.8104}  \newline{}{0.8115}  \newline{}\underline{0.8152} \newline{}\textbf{0.8157} &
		23.52 \newline{}25.84 \newline{}25.70 \newline{}22.04 \newline{}26.61 \newline{}26.77 \newline{}27.50 \newline{}28.46 \newline{}28.81 \newline{}28.48 \newline{}28.83 \newline{}{28.99} \newline{}{28.89} \newline{}{29.17}  \newline{}\underline{29.72} \newline{}\textbf{29.86}  &
		0.6862 \newline{}0.7856 \newline{}0.7770 \newline{}0.6745 \newline{}0.8136 \newline{}0.8154 \newline{}0.8370 \newline{}0.8582 \newline{}0.8647 \newline{}0.8581 \newline{}0.8646 \newline{}{0.8676} \newline{}{0.8661} \newline{}{0.8706}  \newline{}\underline{0.8792} \newline{}\textbf{0.8812}  &
		25.46 \newline{}29.64 \newline{}29.47 \newline{}23.04 \newline{}31.06 \newline{}31.15 \newline{}32.97 \newline{}33.97 \newline{}34.38 \newline{}34.07 \newline{}34.46 \newline{}{34.56} \newline{}{34.49}  \newline{}{34.97}  \newline{}\underline{35.53} \newline{}\textbf{35.66} &
		0.8149\newline{} 0.9003\newline{} 0.8924\newline{} 0.7927\newline{} 0.9234\newline{} 0.9245\newline{} 0.9391\newline{} 0.9465\newline{} 0.9483 \newline{} 0.9466 \newline{} 0.9487 \newline{}{0.9494} \newline{}{0.9492}  \newline{}{0.9512}  \newline{}\underline{0.9538} \newline{}\textbf{0.9543}  \\
		\hline
	\end{tabular}%
	\label{tab-BD}%
\end{table}%
\begin{figure*}[t]
\scriptsize
\centering
\begin{tabular}{cc}
\hspace{-0.4cm}
\begin{adjustbox}{valign=t}
	\begin{tabular}{c}
		\includegraphics[width=0.2\textwidth]{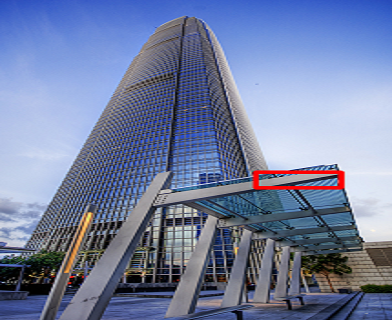}
		\\
		Urban100: img\_046 ($\times$3)
	\end{tabular}
\end{adjustbox}
\hspace{-0.46cm}
\begin{adjustbox}{valign=t}
	\begin{tabular}{cccccc}
		\includegraphics[width=0.2\textwidth]{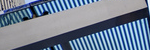} \hspace{-4mm} &
		\includegraphics[width=0.2\textwidth]{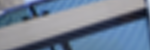} \hspace{-4mm} &
		\includegraphics[width=0.2\textwidth]{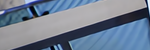} \hspace{-4mm} &
		\includegraphics[width=0.2\textwidth]{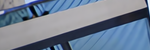} \hspace{-4mm}
		\\
		HR \hspace{-4mm} &
		Bicubic \hspace{-4mm} &
		RDN~\cite{zhang2018residual} \hspace{-4mm} &
		RCAN~\cite{zhang2018image} \hspace{-4mm}
		\\
		\includegraphics[width=0.2\textwidth]{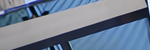} \hspace{-4mm} &
		\includegraphics[width=0.2\textwidth]{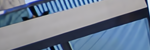} \hspace{-4mm} &
		\includegraphics[width=0.2\textwidth]{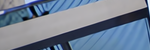} \hspace{-4mm} &
		\includegraphics[width=0.2\textwidth]{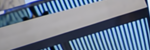} \hspace{-4mm} 
		\\ 
		RFANet~\cite{RFANet}  \hspace{-4mm} &
		DRLN ~\cite{anwar2019drln}  \hspace{-4mm} &
		HAN ~\cite{HAN}  \hspace{-4mm} &
		\model (ours) \hspace{-4mm}
		\\
	\end{tabular}
\end{adjustbox}
\vspace{1mm}
\\

\hspace{-0.4cm}
\begin{adjustbox}{valign=t}
	\begin{tabular}{c}
		\includegraphics[width=0.2\textwidth]{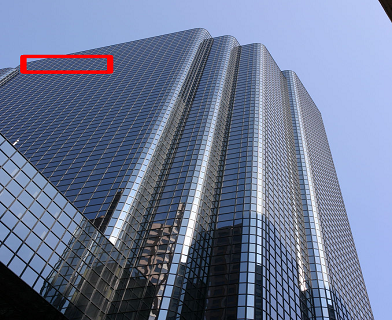}
		\\
		Urban100: img\_074 ($\times$3)
	\end{tabular}
\end{adjustbox}
\hspace{-0.46cm}
\begin{adjustbox}{valign=t}
	\begin{tabular}{cccccc}
		\includegraphics[width=0.2\textwidth]{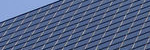} \hspace{-4mm} &
		\includegraphics[width=0.2\textwidth]{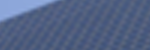} \hspace{-4mm} &
		\includegraphics[width=0.2\textwidth]{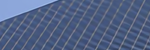} \hspace{-4mm} &
		\includegraphics[width=0.2\textwidth]{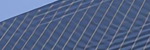} \hspace{-4mm}
		\\
		HR \hspace{-4mm} &
		Bicubic \hspace{-4mm} &
		RDN~\cite{zhang2018residual} \hspace{-4mm} &
		RCAN~\cite{zhang2018image} \hspace{-4mm}
		\\
		\includegraphics[width=0.2\textwidth]{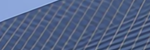} \hspace{-4mm} &
		\includegraphics[width=0.2\textwidth]{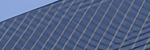} \hspace{-4mm} &
		\includegraphics[width=0.2\textwidth]{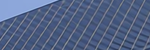} \hspace{-4mm} &
		\includegraphics[width=0.2\textwidth]{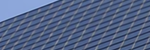} \hspace{-4mm} 
		\\ 
		RFANet~\cite{RFANet}  \hspace{-4mm} &
		DRLN ~\cite{anwar2019drln}  \hspace{-4mm} &
		HAN ~\cite{HAN}  \hspace{-4mm} &
		\model (ours) \hspace{-4mm}
		\\
	\end{tabular}
\end{adjustbox}
\vspace{1mm}
\\

\end{tabular}
\caption{Visual comparison for  $3\times$ SR with BD model on Urban100 dataset.  
}
\label{fig:visual_result_SRBDX3}
\end{figure*}

\subsection{Results with Bicubic Downscale-Noise (DN) Degradation Model}
We apply our method to super-resolve images with the downscale-noise (DN) degradation model, which is widely used in recent SISR papers \cite{dong2014learning,kim2016accurate,dong2016accelerating,zhang2017learning,zhang2017learning,zhang2018residual,CRAN}. For DN degradation, HR image is first downscaled with scaling factor $\times 3$, after which Gaussian noise with noise level 30 is added to it.

\textbf{Quantitative results.} 
In Table \ref{tab-DN} we compare the proposed \model{} model with the following super-resolution methods:
SRCNN~\cite{dong2014learning}, FSRCNN~\cite{dong2016accelerating}, VDSR~\cite{kim2016accurate}, IRCNN\_G~\cite{zhang2017learning}, IRCNN\_C~\cite{zhang2017learning}, RDN~\cite{zhang2018residual},  CRAN~\cite{CRAN}. As shown, our solution achieves consistently better performance than the other methods even without self-ensemble (i.e. \model+). 
For DN degradation we do not provide visual comparison, because the results of recent work is not available publicly.

\begin{table}[t]
	\centering
	\scriptsize
	\caption{Quantitative results with DN degradation model. The best and second best results are highlighted in \textbf{bold} and \underline{underlined}}
	\begin{tabular}{|p{6.5em}|p{2.5em}|p{2.5em}|p{3em}|p{2.5em}|p{3em}|p{2.5em}|p{3em}|p{2.5em}|p{3em}|p{2.5em}|p{3em}|}
		
		\hline
		\multicolumn{1}{|c|}{\multirow{2}{*}{{ Method} }} & \multicolumn{1}{c|}{\multirow{2}{*}{Scale}} & \multicolumn{2}{c}{Set5} & \multicolumn{2}{c}{Set14} & \multicolumn{2}{c}{B100} & \multicolumn{2}{c}{Urban100} & \multicolumn{2}{c|}{ Manga109} \\
		\cline{3-12}  &  &  PSNR  & SSIM  & PSNR   & SSIM  & PSNR  & SSIM  & PSNR  & SSIM  & PSNR  & SSIM \\
		\hline
		
		Bicubic \newline{}SRCNN~\cite{dong2014learning}  \newline{}FSRCNN~\cite{dong2016accelerating} \newline{}VDSR~\cite{kim2016accurate} \newline{}IRCNN\_G~\cite{zhang2017learning}   \newline{}IRCNN\_C~\cite{zhang2017learning} \newline{}RDN~\cite{zhang2018residual} \newline{} CRAN~\cite{CRAN} \newline{} \model (ours) \newline{} \model +(ours)	& 
		$\times3$ \newline{}$\times3$ \newline{}$\times3$ \newline{}$\times3$ \newline{}$\times3$ \newline{}$\times3$ \newline{}$\times3$ \newline{}$\times3$ \newline{}$\times3$ \newline{} $\times3$  &
		24.01 \newline{} 25.01 \newline{}24.18 \newline{}25.20 \newline{}25.70 \newline{}27.48 \newline{}28.47 \newline{}28.74 \newline{}\underline{28.81} \newline{}\textbf{28.84}   &
		0.5369 \newline{}0.6950 \newline{}0.6932 \newline{}0.7183 \newline{}0.7379 \newline{}0.7925 \newline{}0.8151 \newline{}0.8235 \newline{}\underline{0.8244} \newline{}\textbf{0.8251}   &
		22.87 \newline{}23.78 \newline{}23.02 \newline{}24.00 \newline{}24.45 \newline{}25.92 \newline{}26.60 \newline{}26.77 \newline{}\underline{26.87} \newline{}\textbf{26.90} &
		0.4724 \newline{}0.5898 \newline{}0.5856 \newline{}0.6112 \newline{}0.6305 \newline{}0.6932 \newline{}0.7101 \newline{}0.7178 \newline{}\underline{0.7201} \newline{}\textbf{0.7205}  &
		22.92 \newline{}23.76 \newline{}23.41 \newline{}24.00 \newline{}24.28 \newline{}25.55 \newline{}25.93 \newline{}26.04 \newline{}\underline{26.11} \newline{}\textbf{26.12} &
		0.4449 \newline{}0.5538 \newline{}0.5556 \newline{}0.5749 \newline{}0.5900 \newline{}0.6481 \newline{}0.6573 \newline{}0.6647 \newline{}\underline{0.6674} \newline{}\textbf{0.6678}&
		21.63 \newline{}21.90 \newline{}21.15 \newline{}22.22 \newline{}22.90 \newline{}23.93 \newline{}24.92 \newline{}25.43 \newline{}\underline{25.73} \newline{}\textbf{25.82}  &
		0.4687 \newline{}0.5737 \newline{}0.5682 \newline{}0.6096 \newline{}0.6429 \newline{}0.6950 \newline{}0.7354 \newline{}0.7566 \newline{}\underline{0.7654} \newline{}\textbf{0.7678}  &
		23.01 \newline{}23.75 \newline{}22.39 \newline{}24.20 \newline{}24.88 \newline{}26.07 \newline{} 28.00  \newline{} 28.44 \newline{}\underline{28.80} \newline{}\textbf{28.88}  &
		0.5381\newline{} 0.7148 \newline{} 0.7111 \newline{} 0.7525\newline{} 0.7765\newline{} 0.8253\newline{} 0.8591 \newline{} 0.8692 \newline{} \underline{0.8739} \newline{} \textbf{0.8750} \\
		\hline
	\end{tabular}%
	\label{tab-DN}%
\end{table}%

\subsection{Model Complexity Analyses}

\begin{table}[!t]\scriptsize
	
	\tabcolsep 10pt
	\caption{Number of parameters, FLOPs (for $3\times 64 \times 64$ input) and performance on Manga109 with upscale factor $\times 4$ (BI model)}
	
	\begin{center}\scriptsize{
			\begin{tabular}{cccc}
				\toprule
				Model 	& Parameters, M & FLOPs, G & PSNR, dB \\
				\midrule				
				SwinIR  & 11.9          & 54 & 32.03  \\ 
				RCAN    & 15.6        & 65 &  31.21\\ 
				SAN   & 15.9        & 67  & 31.18\\ 
				HAN     & 16.9        & 67  & 31.42\\ 
				$\text{RDRN}_4$ (ours)  & 18.5        & 62 &  32.09\\ 
				\midrule
				NLSN*   & 44.2        & 222 & 32.09\\ 
				NLSN    & 44.2        & 222 & 31.27\\ 
				$\text{RDRN}_5$ (ours)  & 36.4        & 123 &  32.27\\ 
				$\text{RDRN}_6$ (ours)  & 72.2        & 241  & 32.39\\
				\bottomrule		  
		\end{tabular}}
		\label{table:kysymys}
	\end{center} 
\end{table}

Fig. \ref{fig:my} and Table \ref{table:kysymys} demonstrate comparison with recent SR works in terms of model size, FLOPs and performance on Manga109 dataset. Compared to NLSN, the number of parameters of our $\text{RDRN}_5$ is reduced by \textbf{18\%.}

\textbf{Recursion Depth Analysis.} We show comparison of the proposed models with different recursion depth $T = 4,5,6$. 
The smaller network $\text{RDRN}_4$ outperforms most of the recent methods on Manga109 dataset with comparable complexity and number of parameters.

\section{Conclusion}

In this paper, we propose a \modelfullname{} (\model) for highly accurate image SR. Specifically, \blockfullname{} (\blockname) allows us to build and train a large and powerful network. To stabilize the training and further improve the quality we apply intermediate supervision (IS) loss function. Training with IS allows the network to learn more informative features for more accurate reconstruction. \model{} achieves superior SISR results under different degradation models, such as bicubic interpolation (BI), blur-downscale (BD) and downscale-noise (DN). Extensive experiments demonstrate that the proposed model outperforms recent state-of-the-art solutions in terms of accuracy and visual quality. The proposed network architecture is general and could be applied to other low-level computer vision tasks. 
The architecture of the core \blockname{} block can be simply described by two schemes: basic block and the recursive block. We implement only simple ideas and believe that the proposed approach could be significantly improved either using manual or automatic search.

\subsubsection{Acknowledgements} Authors would like to thank Ivan Mazurenko, Li Jieming and Liao Guiming from Huawei  for fruitful discussions, guidance and support.

%
%
%
%
\bibliographystyle{splncs04}
\bibliography{sr2021}

\end{document}


\title{\model: \modelfullnametitle{}  for Image Super-Resolution \newline Supplementary Material}

\titlerunning{\model: \modelfullnametitle{}  for Image Super-Resolution}
%
\author{Alexander Panaetov \inst{1}\orcidID{0000-0003-2309-9798} \and
Karim Elhadji~Daou\inst{1}\orcidID{0000-0003-4677-7571} \and
Igor Samenko\inst{1}\orcidID{0000-0001-9400-312X} \and
Evgeny Tetin\inst{1}\orcidID{ 0000-0001-6878-8330} \and
Ilya Ivanov\inst{1}\orcidID{0000-0001-7919-5143}} 

\authorrunning{A. Panaetov et al.}
%
\institute{Huawei, Moscow Research Center, Russia \email{\{panaetov.alexander1, karim.daou, samenko.igor, evgeny.tetin, ivanov.ilya1\}@huawei.com}}

\maketitle              

\section{Results on Image SR ($\times$ 8)}
We show the comparison on classical image SR ($\times$ 8, BI degradation) in Table  \ref{table-BIX8}.

\begin{table}[!ht]
	\scriptsize 
	\centering
	\caption{Quantitative results with BI degradation model. The best and second best results are highlighted in \textbf{bold} and \underline{underlined}}
\begin{tabular}{|p{7em}|p{2.5em}|p{2.5em}|p{3em}|p{2.5em}|p{3em}|p{2.5em}|p{3em}|p{2.5em}|p{3em}|p{2.5em}|p{3em}|} 
		\hline
		\multicolumn{1}{|c|}{\multirow{2}{*}{{ Methods} }} & \multicolumn{1}{c|}{\multirow{2}{*}{Scale}} & \multicolumn{2}{c}{Set5} & \multicolumn{2}{c}{Set14} & \multicolumn{2}{c}{B100} & \multicolumn{2}{c}{Urban100} & \multicolumn{2}{c|}{ Manga109} \\
		\cline{3-12}   \multicolumn{1}{|c|}{} & \multicolumn{1}{c|}{} &  PSNR  & SSIM  & PSNR   & SSIM  & PSNR  & SSIM  & PSNR  & SSIM  & PSNR  & SSIM \\
\hline
	\end{tabular}%
	\label{table-BIX8}%
\end{table}

\section{Additional Visual Comparison}
We provide more visual comparison for $\times2$, $\times3$, $\times4$ SR with BI degradation model in Figures \ref{fig:visual_result_SRBIX2}, \ref{fig:visual_result_SRBIX3}, \ref{fig:visual_result_SRBIX4} and for $\times3$ SR with BD degradation model in Figure \ref{fig:visual_result_SRBDX3}.

\begin{figure*}[t]
	\scriptsize
	\centering
	%
	\caption{Visual comparison for $2\times$ SR with BI model.}
	\label{fig:visual_result_SRBIX2}
\end{figure*}


\begin{figure*}[t]
	\scriptsize
	\centering
	%
	\caption{Visual comparison for $3\times$ SR with BI model.  
	}
	\label{fig:visual_result_SRBIX3}
\end{figure*}
\begin{figure*}[t]
	\scriptsize
	\centering
	%
\caption{Visual comparison for $4\times$ SR with BI model.  
}
\label{fig:visual_result_SRBIX4}
\end{figure*}

\begin{figure*}[t]
\scriptsize
\centering
%
\end{adjustbox}
\vspace{1mm}
\\
\end{tabular}
\caption{Visual comparison for  $3\times$ SR with BD model.  
}
\label{fig:visual_result_SRBDX3}
\end{figure*}

\section{\blockname{} Implementation Details}
To avoid any misunderstanding about the definition of the proposed \blockname{}, we provide PyTorch implementation of the basic block and the recursive step.
\newpage
\noindent
{\it PyTorch Implementation of \blockfullname}
	\begin{verbatim}

class ConvAdaESA(nn.Module): # RDRB_0
    def __init__(self, channels):
        self.conv = conv(channels, channels, 3)
        self.esa  = ESA(channels)
        self.act  = activation('lrelu')
        self.norm = BatchNorm2d(channels)
        self.phi  = conv(1, 1, 1)
        self.level = 0

    def forward(self, x):
        s   = torch.std(x, dim=[1,2,3], keepdim=True)
        out = self.act(self.conv(self.norm(x)))       # BN + Conv + Act
        out = out * torch.exp(self.phi(torch.log(s))) # AdaDM
        out = self.esa(out+x)                         # Skip + ESA
        return out

def build_RDRB(block):
    class RDRB_base(nn.Module):
        def __init__(self, channels):
            self.block1 = block(channels)
            self.block2 = block(channels)
            self.conv   = conv(channels*2, channels, 1)
            self.esa    = ESA(channels)
            self.act    = activation('lrelu')
            self.level  = self.block1.level + 1
            if self.level == 3:
                self.NLSA = NonLocalSparseAttention(channels=channels)

        def forward(self, x):
            out1 = self.block1(x)
            out2 = self.block2(out1)
            out  = torch.cat([out1,out2],dim=1)
            out  = self.conv(out) + x
            out  = self.esa(self.act(out))
            if self.level == 3:
                out = self.NLSA(out)
            return out
    return RDRB_base


	\end{verbatim}
	\noindent
\noindent

\section{Results on DIV2K validation dataset}

We show results for classical SR on validation DIV2K dataset in Table \ref{table_div2k}.  We employ peak signal-to-noise ratio (PSNR) and structural similarity (SSIM) \cite{SSIM} to measure the quality of super-resolved images. All SR results are evaluated on RGB channels and on Y channel after color space transformation from RGB to YCbCr.
\begin{table}[]
	\begin{tabular}{|l|llll|llll|}
		\hline
		Model                                         & \multicolumn{4}{c|}{RDRN}                                                                      & \multicolumn{4}{c|}{RDRN+}                                                                     \\ \hline
		\multicolumn{1}{|c|}{\multirow{2}{*}{Metric}} & \multicolumn{2}{c|}{Y channel}                           & \multicolumn{2}{c|}{RGB channels}   & \multicolumn{2}{c|}{Y channel}                           & \multicolumn{2}{c|}{RGB channels}   \\ \cline{2-9} 
		\multicolumn{1}{|c|}{}                        & \multicolumn{1}{l|}{PSNR}  & \multicolumn{1}{l|}{SSIM}   & \multicolumn{1}{l|}{PSNR}  & SSIM   & \multicolumn{1}{l|}{PSNR}  & \multicolumn{1}{l|}{SSIM}   & \multicolumn{1}{l|}{PSNR}  & SSIM   \\ \hline
		x2                                            & \multicolumn{1}{l|}{36.98} & \multicolumn{1}{l|}{0.9512} & \multicolumn{1}{l|}{35.47} & 0.9437 & \multicolumn{1}{l|}{37.03} & \multicolumn{1}{l|}{0.9515} & \multicolumn{1}{l|}{35.52} & 0.9440 \\ \hline
		x3                                            & \multicolumn{1}{l|}{33.16} & \multicolumn{1}{l|}{0.8985} & \multicolumn{1}{l|}{31.70} & 0.8855 & \multicolumn{1}{l|}{33.22} & \multicolumn{1}{l|}{0.8992} & \multicolumn{1}{l|}{31.76} & 0.8864 \\ \hline
		x4                                            & \multicolumn{1}{l|}{31.11} & \multicolumn{1}{l|}{0.8515} & \multicolumn{1}{l|}{29.66} & 0.8344 & \multicolumn{1}{l|}{31.17} & \multicolumn{1}{l|}{0.8525} & \multicolumn{1}{l|}{29.73} & 0.8354 \\ \hline
		x8                                            & \multicolumn{1}{l|}{27.32} & \multicolumn{1}{l|}{0.7333} & \multicolumn{1}{l|}{25.89} & 0.7059 & \multicolumn{1}{l|}{27.42} & \multicolumn{1}{l|}{0.7356} & \multicolumn{1}{l|}{25.99} & 0.7083 \\ \hline
	\end{tabular}
	\caption{Quantitative results of SR with BI degradation model on DIV2K validation dataset.}
	\label{table_div2k}
\end{table}

%
%
%
%
\bibliographystyle{splncs04}
\bibliography{sr2021}